\documentstyle[11pt,aaspp4,tighten]{article}

\slugcomment{draft of \today}

\begin{document}

\font\twelvei = cmmi10 scaled\magstep1 
       \font\teni = cmmi10 \font\seveni = cmmi7
\font\mbf = cmmib10 scaled\magstep1
       \font\mbfs = cmmib10 \font\mbfss = cmmib10 scaled 833
\font\msybf = cmbsy10 scaled\magstep1
       \font\msybfs = cmbsy10 \font\msybfss = cmbsy10 scaled 833
\textfont1 = \twelvei
       \scriptfont1 = \twelvei \scriptscriptfont1 = \teni
       \def\mit{\fam1 }
\textfont9 = \mbf
       \scriptfont9 = \mbfs \scriptscriptfont9 = \mbfss
       \def\bmit{\fam9 }
\textfont10 = \msybf
       \scriptfont10 = \msybfs \scriptscriptfont10 = \msybfss
       \def\bmsy{\fam10 }

\def\etal{{\it et al.~}}
\def\eg{{\it e.g.,~}}
\def\ie{{\it i.e.,}}

\title{A Divergence-Free Upwind Code\\
for Multidimensional Magnetohydrodynamic Flows\altaffilmark{4}}
 
\author{Dongsu Ryu\altaffilmark{1},
        Francesco Miniati\altaffilmark{2},
        T. W. Jones\altaffilmark{2},
        and Adam Frank\altaffilmark{3}}

\altaffiltext{1}
{Department of Astronomy \& Space Science, Chungnam National University,
Daejeon 305-764, Korea: ryu@canopus.chungnam.ac.kr}
\altaffiltext{2}
{Department of Astronomy, University of Minnesota, Minneapolis, MN 55455:\\
min@msi.umn.edu, twj@msi.umn.edu}
\altaffiltext{3}
{Department of Physics and Astronomy, University of Rochester, Rochester
NY 14627:\\afrank@alethea.pas.rochester.edu}
\altaffiltext{4}
{Submitted to the Astrophysical Journal}

\begin{abstract}

A description is given for preserving ${\bmsy\nabla}\cdot{\vec B}=0$
in a magnetohydrodynamic (MHD) code that employs the upwind, Total
Variation Diminishing (TVD) scheme and the Strang-type operator
splitting for multi-dimensionality.
The method is based on the staggered mesh technique to constrain the
transport of magnetic field: the magnetic field components are defined
at grid interfaces with their advective fluxes on grid edges,
while other quantities are defined at grid centers.
The magnetic field at grid centers for the upwind step is calculated
by interpolating the values from grid interfaces.
The advective fluxes on grid edges for the magnetic field evolution
are calculated from the upwind fluxes at grid interfaces.
Then, the magnetic field can be maintained with 
${\bmsy\nabla}\cdot{\vec B}=0$ exactly, if this is so initially,
while the upwind scheme is used for the update of fluid quantities.
The correctness of the code is demonstrated through tests 
comparing numerical solutions either with analytic solutions or with
numerical solutions from the code using an explicit divergence-cleaning
method.
Also the robustness is shown through tests involving realistic
astrophysical problems.

\end{abstract}


\clearpage

\section{INTRODUCTION}

The current bloom in computational astrophysics has been fed not
only by dramatic advances in computer hardware, but also by
comparable developments in improved algorithms.
Nowhere has that been more important than in methods to solve
the equations of compressible magnetohydrodynamics (MHD). That system is
the most applicable to describe a vast array of central
astrophysical problems. But, MHD has presented a special
challenge, because of the complexity presented by
three non-isotropically propagating wave families with wide ranging
relative characteristic speeds, and the associated need to solve 
a reduced set of Maxwell's equations along with the equations of 
compressible continuum fluid dynamics. 

The condition ${\bmsy\nabla}\cdot{\vec B}=0$ is a necessary initial
constraint in multi-dimensional MHD flows and should
be preserved during their evolution.
While the differential magnetic induction equation formally 
assumes ${\bmsy\nabla}\cdot
{\vec B}=0$, non-zero ${\bmsy\nabla}\cdot{\vec B}$ can be induced over
time in numerical simulations by numerical errors due to discretization
and operator splitting.
This is because, even though conventional numerical schemes may be
exactly conservative of the advective fluxes in the induction equation,
nothing maintains the magnetic fluxes in the sense of Gauss's Law.
That is, nothing forces conservation of zero magnetic charge within a
finite cell during a time step.
Numerical non-zero ${\bmsy\nabla}\cdot{\vec B}$ usually grows exponentially,
causing an anomalous force parallel to the magnetic field and destroying
the correct dynamics of flows, as pointed by Brackbill \& Barnes (1980).
Those authors, along with Zachary \etal (1994) show that the use
of a modified, non-conservative form of the momentum equation can keep the
non-zero ${\bmsy\nabla}\cdot{\vec B}$ small enough that no further
correction is necessary for this purpose.
However, the modified form is not suitable for some schemes, and more
importantly may result in unphysical results due the non-conservation
of momentum (see, \eg LeVeque 1997).

Several methods have been suggested and used to maintain
${\bmsy\nabla}\cdot{\vec B}=0$ in MHD codes. We mention four here.
In the first method, vector potential is used instead of magnetic field
in the induction equation (see \eg Clarke \etal 1986; Lind \etal 1989).
Although ${\bmsy\nabla}\cdot{\vec B}=0$ is ensured through the
combination of divergence and curl operations, the method results in the
second-order derivatives of the vector potential in the Lorentz force
term of the momentum equation.
So, in order to keep second-order accuracy, for instance, the use of a
third-order scheme for spatial derivatives is required (for detailed
discussion, see Evans \& Hawley 1988 and references therein).
In the second method, the MHD equations are modified by adding source terms
and any non-zero ${\bmsy\nabla}\cdot{\vec B}$ is advected
away from the dynamical region (see Powell 1994 for details).
That method works well for some problems with open boundaries but not in
others, including those with periodic boundaries.
In the third method, an explicit divergence-cleaning scheme is
added as a correction after the step to update fluid quantities (see \eg
Zachary \etal 1994; Ryu \etal 1995a; Ryu \etal 1995b).
The method works well if boundary effects are negligible
or the computational domain is periodic and if the grid used is
more or less regular.
Otherwise, however, the scheme is not easily adaptable.
In the fourth method, the transport of magnetic field is constrained
by the use of a staggered mesh: some quantities including magnetic field
components are defined on grid interfaces while other quantities are defined
at grid centers.
The method has been successfully implemented in schemes based on
an artificial viscosity (see \eg Evans \& Hawley 1988; DeVore 1991; Stone
\& Norman 1992).
However, since it is ``unnatural'' to stagger fluid quantities in
Riemann-solver-based schemes, that approach has only recently
been applied successfully in such schemes.

Conservative, Riemann-solver-based schemes, which are inherently
upwind, have proven to be very
effective for solving MHD equations as well as hydrodynamic equations.
These schemes conservatively update the zone-averaged or grid-centered
fluid and magnetic
field states based on estimated advective fluxes of mass, momentum,
energy  and magnetic
field at grid interfaces using solutions to the Riemann problem
at each interface. 
MHD examples include Brio \& Wu (1988), Zachary \& Colella (1992),
Zachary, \etal (1994), Dai \& Woodward (1994a,1994b), Powell (1994),
Ryu \& Jones (1995), Ryu, \etal (1995a), Powell \etal (1995),
Roe \& Balsara (1996), Balsara (1997), and Kim \etal (1998).
Brio \& Wu applied Roe's approach to the MHD equations.
Zachary and collaborators used the BCT scheme to estimate fluxes,
while Dai \& Woodward applied the PPM scheme to MHD.
Ryu and collaborators extended Harten's TVD scheme to MHD.
Powell and collaborators developed a Roe-type Riemann solver with
an eight-wave structure for MHD (one more than the usual number
of characteristic MHD waves), one of which is used to remove
non-zero ${\bmsy\nabla}\cdot{\vec B}$.
Balsara used also the TVD scheme to build an MHD code.

The upwind schemes share an ability to sharply and cleanly define fluid
discontinuities, especially shocks, and exhibit a robustness that  
makes them broadly applicable.
But, due to the feature of the upwind schemes that zone-averaged or
grid-centered quantities are used to estimate fluxes at grid interfaces,
the staggered mesh technique has been slow to be incorporated for
magnetic flux conservation.
Instead, the explicit divergence-cleaning scheme has been more commonly
used.
However, recently Dai \& Woodward (1998) suggested an approach to
incorporate the staggered mesh technique in the upwind schemes.
It relies on the separation of the update of magnetic field from
that of other quantities.
Quantities other than magnetic field are updated in the upwind step
either by a split or an un-split method.
Then the magnetic field update is done through an un-split operation
after the upwind dynamical step.
Magnetic field components are defined on grid interfaces, and their advective
fluxes are calculated on grid edges using the time-averaged magnetic field
components at grid interfaces and the other time-averaged quantities at
grid centers through a simple spatial averaging.
For the upwind dynamical step, the values of magnetic field at grid centers
are interpolated from those on grid interfaces.

In this paper, we describe an implementation of the Dai \& Woodward
approach into a previously published upwind MHD code based on the 
Harten's Total Variation
Diminishing (TVD) scheme (Harten 1983) which the present authors developed
(Ryu \& Jones 1995; Ryu, \etal 1995a; Ryu, \etal 1995b; Kim \etal 1998).
The TVD scheme is a second-order-accurate extension of the Roe-type
upwind scheme.
The previous code employed an explicit divergence-cleaning technique
in multi-dimensional versions, and has been applied to a variety of
astrophysical problems including the MHD Kelvin-Helmholtz instability
(Frank \etal 1996; Jones \etal 1997), the propagation of supersonic
clouds (Jones \etal 1996), and MHD jets (Frank \etal 1998).
However, the range of application for the code has been limited due
to the restrictions on boundary condition and grid structures, as
pointed out above.
In this paper we address those limitations by incorporating the
staggered mesh algorithm to keep ${\bmsy\nabla}\cdot{\vec B}=0$.
{\it However, instead of calculating the advective fluxes for the magnetic
field update as Dai \& Woodward (1998), we calculate the fluxes at grid
edges using the fluxes at grid interfaces from the upwind step.}
The advantage of our implementation is that the calculated advective
fluxes keep the upwindness in a more obvious way.
We show that our new code performs at least as well as the previous
version in direct comparisons, but also that it effectively handles
problems that could not be addressed with the original code.
We intend this paper to serve as a reference for works which use the
code for astrophysical applications.
In \S2, the numerical method is described, while several tests are
presented in \S3.
A brief discussion follows in \S4.

\section{IMPLEMENTATION OF THE DIVERGENCE-FREE STEP}

We describe the magnetic field update step in two-dimensional plane-parallel geometry.
Extensions to three-dimension and other geometries are trivial.
The induction equation in the limit of negligible electrical resistivity
is written in conservative form as
\begin{equation}
\frac{\partial B_x}{\partial t} + \frac{\partial}{\partial y}
\left(B_x v_y - B_y v_x\right) = 0,
\end{equation}
and
\begin{equation}
\frac{\partial B_y}{\partial t} + \frac{\partial}{\partial x}
\left(B_y v_x - B_x v_y\right) = 0.
\end{equation}
The full MHD equations in conservative form can be found, for instance,
in Ryu \etal (1995a).
In typical upwind schemes, including the TVD scheme for MHD equations
as well as hydrodynamic equations, fluid quantities are zone-averages
or defined at
grid centers. Their advective fluxes are calculated on grid interfaces
through approximate solutions to the Riemann problem there, based
on interpolated values.

Here, we define the magnetic field components on {\it grid interfaces},
$b_{x,i,j}$ and $b_{y,i,j}$, while all the other fluid quantities are
still defined at {\it grid centers} (See Fig. 1 for the notations used in this
paper).
For use in the step of calculating the advective fluxes by the TVD scheme,
the magnetic field components at grid centers, which are intermediate
variables, are interpolated as
\begin{equation}
B_{x,i,j} = \frac{1}{2}\left(b_{x,i,j}+b_{x,i-1,j}\right),
\end{equation}
and
\begin{equation}
B_{y,i,j} = \frac{1}{2}\left(b_{y,i,j}+b_{y,i,j-1}\right).
\end{equation}
Since the MHD code based on the TVD scheme has second-order accuracy,
the above second-order interpolation should be adequate.
If non-uniform grids are used, an appropriate interpolation of
second-order should be used.
With the fluid quantities, including these magnetic field values,
given at grid centers,
the TVD advective fluxes are used to update the fluid quantities
from the time step $n$ to $n+1$
\begin{equation}
{\vec q}_{i,j}^{n+1}=L_yL_x{\vec q}_{i,j}^n,
\end{equation}
as described in Ryu \& Jones (1995) and Ryu \etal (1995a).
Strang-type operator splitting is used, so that the operation $L_xL_y$
is applied from $n+1$ to $n+2$.
Here, ${\vec q}$ is the state vector of fluid variables.

The advective fluxes used to update the magnetic field components at
grid interfaces are calculated also during the TVD step from the MHD
Riemann solution with little additional
cost.
In the $x$-path the following flux is computed
\begin{equation}
{\bar f}_{x,i,j} = \frac{1}{2}\left(B_{y,i,j}^n v_{x,i,j}^n
+ B_{y,i+1,j}^n v_{x,i+1,j}^n\right) - \frac{\Delta x}{2\Delta t^n}
\sum_{k=1}^7 \beta_{k,i+\frac{1}{2},j}^n R_{k,i+\frac{1}{2},j}^n(5),
\end{equation}
and in the $y$-path the following flux is computed
\begin{equation}
{\bar f}_{y,i,j} = \frac{1}{2}\left(B_{x,i,j}^n v_{y,i,j}^n
+ B_{x,i,j+1}^n v_{y,i,j+1}^n\right) - \frac{\Delta y}{2\Delta t^n}
\sum_{k=1}^7 \beta_{k,i,j+\frac{1}{2}}^n R_{k,i,j+\frac{1}{2}}^n(5).
\end{equation}
Here, $\beta_{k,i+\frac{1}{2},j}$ and $\beta_{k,i,j+\frac{1}{2}}$
are the quantities computed at grid interfaces.
As described in Ryu \& Jones (1995),
$\beta_{k,i+\frac{1}{2},j}$ used in the the $x$-path is calculated
as follows:
\begin{equation}
\beta_{k,i+\frac{1}{2},j} = Q_k\left({\Delta t^n\over\Delta x}
a_{k,i+\frac{1}{2},j}^n+\gamma_{k,i+\frac{1}{2},j}\right)
\alpha_{k,i+\frac{1}{2},j}-(g_{k,i,j}+g_{k,i+1,j}),
\end{equation}
\begin{equation}
\alpha_{k,i+\frac{1}{2},j} = {\vec L}_{k,i+\frac{1}{2},j}^n\cdot
({\vec q}_{i+1,j}^n-{\vec q}_{i,j}^n),
\end{equation}
{\textfont1 = \twelvei
       \scriptfont1 = \twelvei \scriptscriptfont1 = \teni
       \def\mit{\fam1 }
\begin{equation}
\gamma_{k,i+\frac{1}{2},j} = \cases{{g_{k,i+1,j}-g_{k,i,j}\over
\alpha_{k,i+\frac{1}{2},j}}_,&for $\alpha_{k,i+\frac{1}{2},j}\neq0$\cr
0, &for $\alpha_{k,i+\frac{1}{2},j}=0$\cr}_,
\end{equation}}
\begin{equation}
g_{k,i,j} = {\rm sign}({\tilde g}_{k,i+\frac{1}{2},j})~
{\rm max}\left[0,~{\rm min}\left\{|{\tilde g}_{k,i+\frac{1}{2},j}|,~
{\tilde g}_{k,i-\frac{1}{2},j}{\rm sign}({\tilde g}_{k,i+\frac{1}{2},j})
\right\}\right],
\end{equation}
\begin{equation}
{\tilde g}_{k,i+\frac{1}{2},j} = \frac{1}{2}
\left[Q_k({\Delta t^n\over\Delta x}a_{k,i+\frac{1}{2},j}^n)
-({\Delta t^n\over\Delta x}a_{k,i+\frac{1}{2},j}^n)^2\right]
\alpha_{k,i+\frac{1}{2},j},
\end{equation}
{\textfont1 = \twelvei
       \scriptfont1 = \twelvei \scriptscriptfont1 = \teni
       \def\mit{\fam1 }
\begin{equation}
Q_k(\chi) = \cases{{\chi^2\over4\varepsilon_k}+\varepsilon_k,
& for $|\chi|<2\varepsilon_k$\cr
|\chi|, & for $|\chi|\geq2\varepsilon_k$\cr}_.
\end{equation}}
$\beta_{k,i,j+\frac{1}{2}}$ used in the $y$-path is calculated
similarly.
$R_{k,i+\frac{1}{2},j}^n(5)$ and $R_{k,i,j+\frac{1}{2}}^n(5)$ are the
fifth components (``$B_y$'' and ``$B_x$'', respectively for the
two passes) of the right eigenvectors and ${\vec L}_k^n$'s are the
left eigenvectors. They are computed on grid interfaces and given
in Ryu \& Jones (1995).
$a_k^n$'s are the speeds of seven characteristic waves which are
also computed on grid interfaces.
Along the $x$-path, they are in non-increasing order
\begin{equation}
a_{1,7} = v_x\pm c_f,\qquad a_{2,6} = v_x\pm c_a,\qquad
a_{3,5} = v_x\pm c_s,\qquad a_4 = v_x.
\end{equation}
$a_k^n$'s along the $y$-path is computed by replacing $v_x$ with $v_y$.
Here, $c_f$, $c_a$, and $c_s$ are local fast, Alfv\'en, and slow speeds,
respectively.
$\varepsilon_k$'s are the internal parameters to control dissipation
in each characteristic wave and should be between 0 and 0.5 (see the
next section).
The time step $\Delta t^n$ is restricted by the usual Courant
condition for the stability.

Using the above fluxes at grid interfaces, the advective fluxes, or the
$z$-component of the electric field, on {\it grid edges} (see Fig. 1 for
definition) are calculated by a simple arithmetic average, which still
keeps second-order accuracy:
namely,
\begin{equation}
{\bar\Omega}_{i,j} = \frac{1}{2}\left({\bar f}_{y,i+1,j}
+{\bar f}_{y,i,j}\right) - \frac{1}{2}\left({\bar f}_{x,i,j+1}
+{\bar f}_{x,i,j}\right).
\end{equation}
Then, the magnetic field components are updated as
\begin{equation}
b_{x,i,j}^{n+1} = b_{x,i,j}^n - \frac{\Delta t^n}{\Delta y}
\left({\bar\Omega}_{i,j}-{\bar\Omega}_{i,j-1}\right),
\end{equation}
and
\begin{equation}
b_{y,i,j}^{n+1} = b_{y,i,j}^n + \frac{\Delta t^n}{\Delta x}
\left({\bar\Omega}_{i,j}-{\bar\Omega}_{i-1,j}\right).
\end{equation}
Note that the $\bar\Omega$ terms include information from seven
characteristic waves.
It is also clear that the net magnetic flux across grid interfaces
is exactly kept to be zero at the step $n+1$
\begin{equation}
\oint_S{\vec b}^{n+1}\cdot d{\vec S} =
(b_{x,i,j}^{n+1}-b_{x,i-1,j}^{n+1})\Delta y
+(b_{y,i,j}^{n+1}-b_{y,i,j-1}^{n+1}){\Delta x}=0,
\end{equation}
if it is zero at the step $n$.

The reason why we take the fluxes in (6) and (7) from the upwind
fluxes for the transport of the magnetic field at grid centers is this.
As can be seen in equation (2) along the $x$-path, it is
$\partial(B_yv_x)/\partial x$ that contains the advective term and
requires modification of fluxes to avoid numerical problems.
$\partial(B_xv_y)/\partial x$ causes no problems.
The same argument is applied to $\partial(B_xv_y)/\partial y$ along
the $y$-path.
We note that in the above scheme, the results of one-dimensional
problems calculated with the two-dimensional code reduce to those
calculated with the one-dimensional code, as it should be.
For instance, in shock tube problems with structures propagating along a
coordinate axis, the two-dimensional code produces exactly the same results as
those given in Ryu \& Jones (1995).
However, with Dai \& Woodward's (1998) advective fluxes, that is not
necessarily true.

The code runs at $\sim400$ Mflops on a Cray C90, similar to the previous
code (Ryu \etal 1995a).
It corresponds to an update rate of $\sim1.2\times10^5$ zones s$^{-1}$
for the two-dimensional version, about a 20\% speed-up compared to the
previous code, due to the absence of the explicit divergence-cleaning step.

\section{NUMERICAL TESTS}

The numerical scheme described in the last section was tested
with two-dimensional problems in plane-parallel and cylindrical geometries
in order to demonstrate its correctness and accuracy as well as to
show its robustness and flexibility.
In all the tests shown, we used the adiabatic index $\gamma=5/3$ and
a Courant constant $C_{cour}=0.8$.
For the internal parameters, $\varepsilon_k$'s, of the TVD scheme
(Ryu \& Jones 1995; Ryu \etal 1995a),
$\varepsilon_{1,7}=0.1-0.2$ (for fast mode),
$\varepsilon_{2,6}=0.05-0.1$ (for slow mode),
$\varepsilon_{3,5}=0-0.05$ (for Alfv\'en mode), and $\varepsilon_4=0-0.1$
(for entropy mode) were used.
However, the test results are mostly not very sensitive to $C_{cour}$ and 
$\varepsilon_k$ values.

\subsection{Shock Tube Problems}

We first tested the code with MHD shock tube problems placed diagonally
on a two-dimensional plane-parallel grid.
The correctness and accuracy are demonstrated through the comparison
of the numerical solutions with the exact analytic solutions from the
non-linear Riemann solver described in Ryu \& Jones (1995).
The calculations were done in a box of
$x=[0,1]$ and $y=[0,1]$, where structures 
propagate along the diagonal line joining $(0,0)$ and $(1,1)$.  
Two examples are presented.
The first, shown in Fig. 2a, includes only two ($x$ and $y$) components
of magnetic field and velocity, so that they are confined in the
computational plane.
The second, shown in Fig. 2b, includes all three vector field components.
The numerical solutions are marked with dots, and the exact analytic solutions
are drawn with lines.
Structures are measured along the diagonal line joining $(0,0)$ and $(1,1)$.
The plotted quantities are density, gas pressure, total energy,
$v_{\|}$ (velocity parallel to the diagonal line; i.e., parallel to
the wave normal), $v_{\bot}$ (velocity perpendicular to the diagonal
line but still in the computational plane), $v_z$ (velocity in
the direction out of plane), and the analogous magnetic field
components, $B_{\|}$, $B_{\bot}$, and $B_z$.

In Fig, 2a, the initial left state is
($\rho$, $v_{\|}$, $v_{\bot}$, $v_z$, $B_{\bot}$, $B_z$, $E$)
= ($1$, $10$, $0$, $0$, $5/\sqrt{4\pi}$, $0$, $20$)
and the initial right state is
($1$, $-10$, $0$, $0$, $5/\sqrt{4\pi}$, $0$, $1$),
with $B_{\|}=5/\sqrt{4\pi}$.
The calculation was done using $256\times256$ cells, and
plots correspond to time $t=0.08\sqrt{2}$.
The structures are bounded by a left and right facing fast shock pair.
There are also a left facing slow rarefaction,
a right facing slow shock and a contact discontinuity.
All are correctly reproduced.
The captured shocks and contact discontinuity here are very similar to
those with the code using an explicit divergence-cleaning scheme,
which was shown in Fig. 1 of Ryu \etal (1995a).

In Fig. 2b, the initial left state is
($\rho$, $v_{\|}$, $v_{\bot}$, $v_z$, $B_{\bot}$, $B_z$, $E$)
= ($1.08$, $1.2$, $0.01$, $0.5$, $3.6/\sqrt{4\pi}$, $2/\sqrt{4\pi}$, $0.95$)
and the initial right state is
($1$, $0$, $0$, $0$, $4/\sqrt{4\pi}$, $2/\sqrt{4\pi}$, $1$),
with $B_{\|}=2/\sqrt{4\pi}$.
Again the calculation was done using $256\times256$ cells, and
plots correspond to time $t=0.2\sqrt{2}$.
Fast shocks, rotational discontinuities, and slow shocks propagate
from each side of the contact discontinuity, all of which are
correctly reproduced.
Again, the structures are captured very similarly to what we found with
the code using an explicit divergence-cleaning scheme shown in Fig. 2
of Ryu \etal (1995a).

\subsection{The Orszag-Tang Vortex}

As a second, and truly multi-dimensional test, we followed the formation 
of the compressible Orszag-Tang vortex.
The problem was originally studied by Orszag \& Tang (1979) in the
context of incompressible MHD turbulence, and later used to test
compressible MHD codes by Zachary, \etal (1994), Ryu \etal (1995a), and Dai \&
Woodward (1998).
Comparison of our new solution with previous ones demonstrates
the correctness of the new code in this problem.

The test was set up in a two-dimensional periodic box of
$x=[0,1]$ and $y=[0,1]$ with $256\times256$ cells.
Initially, velocity is given as
${\vec v} = v_o\left[-\sin(2\pi y){\hat x}+\sin(2\pi x)
{\hat y}\right]$
and magnetic field as
${\vec B} = B_o\left[-\sin(2\pi y){\hat x}+\sin(4\pi x)
{\hat y}\right]$
with $v_o=1$ and $B_o=1/\sqrt{4\pi}$.
Uniform background density and pressure were assumed
with values fixed by
$M^2\equiv{v_o / (\gamma p_o/\rho_o)}=1$,
$\beta\equiv{p_o / (B_o^2/2)}={10/3}$ and $\gamma=5/3$.

Fig. 3 shows the gray scale images of gas pressure, magnetic pressure,
compression, ${\bmsy\nabla}\cdot{\vec v}$, and vorticity,
$\left({\bmsy\nabla}\times {\vec v}\right)_z$ at time $t=0.48$,
as well as the line cuts of gas pressure and magnetic pressure through
$y=0.4277$.
The structures, including fine details, match exactly with those in
Ryu \etal (1995a), proving the correctness of our code.
Although only approximately the same initial conditions and epoch
as those in Zachary, Malagoli, \& Colella (1994) and
Dai \& Woodward (1998) are used, the images show that the overall 
shape and dynamics match
closely with those solutions, as well.

\subsection{Propagation of a Supersonic Cloud}

As an initial test for the robustness and flexibility of the code in a
practical, astrophysical application, we 
simulated supersonic cloud propagation through a magnetized medium.
We present three simulations differing in initial
magnetic field orientation. The first two reproduce 
Models A2 and T2 from Jones \etal (1996), in order to compare with
previous calculations.
The third is a new simulation.
All three were computed on a Cartesian grid.
In the first two cases the magnetic field is successively parallel to the 
cloud motion (aligned case - A2) and perpendicular to it
(transverse case - T2).
In the third, we present a new case with the magnetic field making an angle 
$\theta=45^\circ$ with the cloud velocity (\ie an oblique case).

For these calculations we used the same physical parameters
as in Jones \etal (1996). 
The cloud is initially in pressure equilibrium with the 
background medium and $p_{o}=1/\gamma$ throughout. In addition,
$\rho_{ambient}=1$, and the cloud density 
$\rho_c=\chi~\rho_{ambient}=10 $. A thin boundary layer with 
hyperbolic tangent density profile and characteristic width of two zones
separates the cloud from the background gas. The background sound speed is 
$a_{ambient}=(\gamma p_{o}/\rho_{ambient})^{1/2}=1$. 
At the outset of each simulation, the ambient gas is set into motion around
the cloud with a Mach number $M=10$.
The magnetic field lies in the computational plane and is
initially uniform throughout it. Its strength corresponds to $\beta_0=4$.
Therefore, it is $(B_x,B_y, B_z)=(0.55,0,0)$ for the aligned case,
$(B_x,B_y,B_z)=(0,0.55,0)$ for the transverse case and 
$(B_x,B_y,B_z)=(0.39,0.39,0)$ for the oblique case.
As in Jones \etal (1996) for the aligned and the transverse 
field cases the computational domain is $[x,y]=[10,5]$, whereas 
for the oblique field case it is $[x,y]=[20,10]$. 
The resolution is always 50 zones per initial cloud radius, $R_{cloud}=1$. 

Images of the aligned and transverse cases
are presented in Fig. 4a. Left panels correspond to the
transverse case, and show density images (top two panels) and magnetic field
lines (bottom panels) for two evolutionary times, namely $t/t_{bc}=2$, $6$,
where $t_{bc}$ is the bullet crushing time
(see Jones \etal 1996 for detail).
These are approximately the same times as those shown in Figs. 1 and 2 in
Jones \etal (1996).
Figures representing the aligned field case are analogously 
illustrated in the right panels.
As we can see there is a general agreement in both the density distribution
and the magnetic field structure between the cases in Fig. 4a and the 
corresponding cases (T2 and A2) in Jones \etal (1996).
Minor differences appear in the details of the cloud shape and
the magnetic field adjacent to the cloud for the aligned field 
case at $t=6t_{bc}$. As 
pointed out in Jones \etal (1996), the nonlinear evolution of
these clouds depends very sensitively on the
exact initial perturbations and their growth. For both sets of
simulations the perturbations develop out of geometrical mismatches
between the cloud and the grid.
We showed in that
paper, for example, that consequently a simple shift of the initial 
cloud center by 0.5 zone on the $x$ axis caused differences much greater 
than those illustrated here.
Similarly, even minor changes in the field adjacent to the cloud near
the start of the calculation or in
the dissipation constants used can be expected to lead to observable
changes in the detailed cloud features.
Thus, by considering different schemes to keep ${\bmsy\nabla}\cdot{\vec B}=0$
used as well as different values of $C_{cour}$ and $\varepsilon_k$'s used in
the different sets of simulations, we judge
the agreement between the two codes to be good.
In addition, the new code seems better able to handle the
extreme rarefaction that forms to the rear of the cloud
immediately after it is set in motion. That is a severe test, since
the plasma $\beta$ abruptly drops from values larger than unity to values
smaller than $\beta \sim 10^{-2}$.

The simulation of a cloud interacting with an oblique magnetic
field offers a good example of the increased flexibility of the
new code.
This situation is more realistic than the other two, but difficult
to simulate with the old code because of its lack of a suitable
periodic space for solving Poisson's equation in the explicit
divergence-cleaning step.
Since the aligned and 
transverse field cases differ considerably in their dynamics, it is 
astrophysically important to be able to investigate the 
general case of an oblique magnetic field.
Fig. 4b illustrates the properties of one such simulation with the new
code.
Further details are discussed in Miniati \etal (1998b).
Top and bottom panels correspond to density distribution and field line
geometry respectively for two different evolutionary times
(again $t/t_{bc}=2$, $6$).
As we can see the evolution of the oblique case produces several features 
analogous to the previous transverse field case.
In particular the magnetic field lines drape around the cloud 
nose and form an intense magnetic region there, due to field 
line stretching. In this fashion the field lines compress 
the already shock crushed cloud and prevent the rapid growth of the
Kelvin-Helmholtz and Rayleigh-Taylor instabilities (Jones \etal 1996). 
However, the broken symmetry across the motion axis also generates uneven 
magnetic field tension that causes some rotation  and lateral
motion of the cloud. In addition, it enhances 
turbulent motions in the wake and, therefore, the onset of the 
tearing mode instability and magnetic reconnection there. 

\subsection{Jets}

As a final test to confirm the robustness of the new code and
to demonstrate its application with a different grid geometry, we
illustrate the simulation of a light cylindrical MHD
jet with a top-hat velocity profile.
The jet enters a cylindrical box of $r=[0,1]$ and $z=[0,6.64]$ at
$z=0$.
The grid of the box is uniform with $256\times1700$ cells and the jet
has a radius, $r_{jet}$, of $30$ cells.
The ambient medium has sound speed $a_{ambient}=1$ and poloidal
magnetic field  ($B_\phi = B_r = 0$, $B_z = B_{ambient}$) with magnetic
pressure 1\% of gas pressure (plasma $\beta_{ambient} = 100$).
The jet has Mach number $M_{jet}\equiv v_{jet}/a_{ambient} =20$, gas
density contrast $\rho_{jet}/\rho_{ambient}=0.1$, and gas pressure in
equilibrium with that of the ambient medium.
It carries a helical magnetic field with $B_r = 0$,
$B_\phi = 2\times B_{ambient} (r/r_{jet})$, and $B_z = B_{ambient}$.
The jet is slightly over-pressured due to the additional $B_\phi$
component, but the additional pressure is too small to have any
significant dynamical consequences.
The simulation was stopped at $t=2.2$ when the bow shock
reached the right boundary.
It takes about about 20 CPU hours on a Cray C90 processor 
or about 90 CPU hours
on a SGI Octane with a 195 MHz R10000.

Fig. 5 shows the images of the log of the gas density and total
magnetic field pressure (Fig. 5a) and the velocity vectors and the $r$ and
$z$ magnetic field vector components (Fig. 5b) at five different epochs,
$t=0.3$, $0.8$, $1.3$ $1.8$, and $2.2$.
The length of velocity arrows is scaled as $\sqrt{|v|}$ and that of
magnetic field arrows as $|B|^{1/4}$.
The figures exhibit clearly the complexity and unsteadiness of the flows.
By viewing an animation of the simulation, which is posted
in the web site http://canopus.chungnam.ac.kr/ryu/testjet/testjet.html,
it becomes obvious that all of the structures are ephemeral and/or highly
variable.
The most noticeable structures are the bow shock of the ambient
medium and the terminal shock of the jet material.
In addition, the jet material expands and then refocuses alternately as it
flows and creates several internal oblique shocks, as described
in many previous works, for example, Lind \etal (1989).
The terminal and oblique shocks are neither steady nor stationary
structures.
The oblique shocks interact episodically with the terminal 
shock, resulting in disruption and reformation of the terminal
shocks.
The terminal shock includes a Mach stem, so that the jet material near
the outside of the jet exits through the oblique portion of the shock.
That material carries vorticity and forms a cocoon around the jet.
The vorticity is further developed into complicated turbulent flows
in the jet boundary layer, which is subject to the Kelvin-Helmholtz
instability.
There are distinct episodes of strong vortex shedding which coincide
with disruption and reformation of the terminal shock.
Its remnants are visible as rolls in the figures.

Although the total magnetic field in the back-flow region is strong
compared to $B_{ambient}$,
as can be seen
in the $P_b$ images, the components
$B_r$ and $B_z$ are comparatively small, as can be seen in the vector plots.
This is because reconnection induced by the complicated turbulent
flow motion of the jet material has frequently annihilated 
$B_r$ and $B_z$, at the same time that $B_\phi$ has been enhanced 
by stretching.
In an axis-symmetric calculation, the $B_\phi$ component cannot
be modified by reconnection, since it is decoupled from the
other two magnetic field components.
We emphasize that the details of the magnetic field configuration
are sensitive to the assumed helical field within the incoming jet, so
that our test results are representative only.

Good agreement of this simulation with previous works such as
Lind \etal (1989) provides another confirmation of validity and
applicability of the new code.
Detailed discussion of comparable jet simulations carried out
with this new code  in the context of radio galaxies, including
acceleration and transport of relativistic electrons,
have been reported in Jones \etal (1998).

\section{DISCUSSION}

For ordinary gasdynamics development
of conservative, high order, monotonicity-preserving, Riemann-solution 
based algorithms, such as the TVD
scheme employed here, provided a key step by enabling stable, accurate
and sharp capture of strong discontinuities expected in compressible
flows, while efficiently following smooth flows with a 
good economy of grid cells. The methods maintain
exact mass, energy and momentum conservation and seem to do a good
job of representing sub-grid-scale dissipation processes (\eg Porter
and Woodward 1994).
Recent extension of those
methods to MHD have also shown great promise, since they
offer the same principal advantages as in gasdynamics. The main disadvantage
of the Riemann methods in MHD were, until now, that they
are basically finite volume schemes, so that they depend on
knowing information averaged over a zone volume, or equivalently
in second order schemes, at grid centers. The problem this presented
came from the fact that the conservation of magnetic charge
depends on a surface integral constraint, which is not guaranteed
by the conservation of advective fluxes used in the remaining
set of MHD relations. As discussed in the introduction this
can lead to physically spurious results. 

Consequently, it is a significant advance to develop accurate,
efficient and robust schemes for maintaining zero magnetic
charge that are adaptable to Riemann-based methods. The method
discussed in this paper seems to be an excellent choice. {\it Since it
exactly conserves the surface integral of magnetic flux over a
cell and does it in an upwind fashion}, it represents a class of
techniques that have come to be called ``Method of Characteristics,
Constrained Transport" or ``MoCCT".
In this paper we outline a specific implementation of this scheme
inside a multi-dimensional MHD extension of Harten's TVD scheme.
With our prescription it should be straightforward for other
workers to accomplish the same outcome. Through a varied bank of 
test problems we have been able to demonstrate the accuracy and
the flexibility of the methods we have employed. Thus, we believe
this code and others like it offer great potential for exploration
of a wide variety of important astrophysical problems. Already
the code described in the paper has been used successfully in
Jones \etal (1998), Miniati \etal (1998a), and Miniati \etal (1998b)
to study propagation of cylindrical MHD jets including the
acceleration and transport of relativistic electrons, and to 
study the propagation and collision between interstellar
plasma clouds.

\acknowledgments

The work by DR was supported in part by KOSEF through the 1997
Korea-US Cooperative Science Program 975-0200-006-2.
The work by TWJ and FM was supported in part by the NSF through grants
AST93-18959, INT95-11654, AST96-19438 and AST96-16964, by NASA grant 
NAG5-5055 and by the University of Minnesota Supercomputing Institute.
The authors are grateful to the referee for clarifying comments.

\clearpage

\figurenum{1}
\figcaption{Notations for the flow variables used in the paper.
The centered magnetic field, $B$, and the velocity, $v$, are defined
at {\it grid centers}.
The faced magnetic field, $b$, and the upwind fluxes, $f$, are defined
on {\it grid interfaces}.
The advective fluxes for the magnetic field update, $\Omega$, is
defined at {\it grid edges}.}

\figurenum{2a}
\figcaption{Two-dimensional MHD shock tube test.
Structures propagate diagonally along the line from $(0,0)$ to $(1,1)$ in
the $x-y$ plane.
The initial left state is
($\rho$, $v_{\|}$, $v_{\bot}$, $v_z$, $B_{\bot}$, $B_z$, $E$)
= ($1$, $10$, $0$, $0$, $5/\sqrt{4\pi}$, $0$, $20$)
and the initial right state is
($1$, $-10$, $0$, $0$, $5/\sqrt{4\pi}$, $0$, $1$)
with $B_{\|}=5/\sqrt{4\pi}$ (same test as Fig. 1 in Ryu \etal 1995a).
The plots are shown at time $t=0.08\sqrt{2}$ along the diagonal line
joining $(0,0)$ and $(1,1)$.
Dots represent the numerical solution from the code described in \S 2
with $256\times256$ cells.
Lines represent the exact analytic solution from the nonlinear Riemann
solver described in Ryu \& Jones (1995).}

\figurenum{2b}
\figcaption{Two-dimensional MHD shock tube test.
Structures propagate diagonally along the line from $(0,0)$ to $(1,1)$ in
the $x-y$ plane.
The initial left state is
($\rho$, $v_{\|}$, $v_{\bot}$, $v_z$, $B_{\bot}$, $B_z$, $E$)
= ($1.08$, $1.2$, $0.01$, $0.5$, $3.6/\sqrt{4\pi}$, $2/\sqrt{4\pi}$, $0.95$)
and the initial right state is
($1$, $0$, $0$, $0$, $4/\sqrt{4\pi}$, $2/\sqrt{4\pi}$, $1$)
with $B_{\|}=2/\sqrt{4\pi}$ (same test as Fig. 2 in Ryu \etal 1995a).
The plots are shown at time $t=0.2\sqrt{2}$ along the diagonal line
joining $(0,0)$ to $(1,1)$.
Dots represent the numerical solution from the code described in \S 2
with $256\times256$ cells.
Lines represent the exact analytic solution from the nonlinear Riemann
solver described in Ryu \& Jones (1995).}

\figurenum{3}
\figcaption{Gray scale images of gas pressure (upper left),
magnetic pressure (upper right), ${\bmsy\nabla}\cdot{\vec v}$
(lower left), and $\left({\bmsy\nabla}\times{\vec v}\right)_z$
(lower right) in the compressible Orszag-Tang vortex test.
White represents high (or positive) values and black represents
low (or negative) values.
The calculation has been done in a periodic box of $x=[0,1]$ and
$y=[0,1]$ with $256\times256$ cells.
The initial configuration is $\rho=25/36\pi$, $p=5/12\pi$,
${\vec v} = -\sin(2\pi y){\hat x}+\sin(2\pi x){\hat y}$,
and ${\vec B} = \left[-\sin(2\pi y){\hat x}+\sin(4\pi x)
{\hat y}\right]/\sqrt{4\pi}$, and the images shown are at $t=0.48$.
The line plots show the profiles of gas pressure and magnetic pressure
along $y=0.4277$.}

\figurenum{4a}
\figcaption{A supersonic cloud moving through a magnetized medium
and computed on a Cartesian grid.
The initial cloud radius in each case is 50 cells.
Upper panels show logarithmic gray scale images of gas density, with 
white referring to high density values and black to low ones.
Lower panels illustrate magnetic field lines obtained as contours of the 
magnetic flux function. The initial magnetic field, which lies within
the computational plane, is perpendicular to the 
cloud motion in the left panels (transverse case) and parallel to it in the 
right panels (aligned case). For each quantity and in each case
two different times; namely $t/t_{bc}=2$, $6$, are shown.}

\figurenum{4b}
\figcaption{Same as in Fig. 4a but now for the oblique field case.
The initial magnetic field makes an angle $\theta=45^\circ$ with 
respect to the cloud motion. The resemblance between
this case with the transverse case is noteworthy.}

\figurenum{5a}
\figcaption{A light MHD cylindrical jet.
The calculation has been done on a $256\times1700$ cylindrical 
grid with a computational domain $r=[0,1]$ and
$z=[0,6.64]$.
The sound speed of the ambient medium, $a_{ambient}=1$, and its magnetic
field is poloidal with $\beta_{ambient}=100$.
The jet has a radius of $30$ cells, density contrast
$\rho_{jet}/\rho_{ambient}=0.1$, and Mach number $M_{jet}=20$.
The jet magnetic field is helical with maximum $\beta_{jet}=20$ at the
surface.
The gray scale images show logarithmic gas density (upper frames)
and logarithmic total magnetic pressure (lower frames) at $t=0.3$,
$0.8$, $1.3$ $1.8$, and $2.2$.
White represents high values and black represents low values.}

\figurenum{5b}
\figcaption{The same jet as in Fig. 5a. The arrows show velocity
(upper frames) and $r$ and $z$-components of magnetic field (lower frames)
at $t=0.3$, $0.8$, $1.3$ $1.8$, and $2.2$.
The length of velocity arrows is scaled as $\sqrt{|v|}$ and that of
magnetic field arrows as $|B|^{1/4}$, in order to clarify the structures
with small velocity and magnetic field magnitudes.}

\end{document}